\documentclass[conference]{IEEEtran}
\IEEEoverridecommandlockouts
\usepackage{cite}
\usepackage{amsmath,amssymb,amsfonts}
\usepackage{algorithmic}
\usepackage{graphicx}
\usepackage{textcomp}
\usepackage{xcolor}
\usepackage{caption}
\usepackage{subcaption}
\usepackage{verbatim}
\usepackage[export]{adjustbox}
\usepackage{xcolor}

\usepackage[colorlinks,citecolor=blue,urlcolor=blue]{hyperref}
\usepackage{cleveref}

\def\BibTeX{{\rm B\kern-.05em{\sc i\kern-.025em b}\kern-.08em    T\kern-.1667em\lower.7ex\hbox{E}\kern-.125emX}}
    
\begin{document}

\title{Enhancing Fidelity of Quantum Cryptography using Maximally Entangled Qubits\\
}
\vspace{-.2in}
%\begin{comment}
\author{\IEEEauthorblockN{Saiful Islam Salim\IEEEauthorrefmark{1},
Adnan Quaium\IEEEauthorrefmark{2}, Sriram Chellappan\IEEEauthorrefmark{3} and A. B. M. Alim Al Islam\IEEEauthorrefmark{4}}
\IEEEauthorblockA{\IEEEauthorrefmark{1}\IEEEauthorrefmark{2}\IEEEauthorrefmark{4}Department of CSE,
Bangladesh University of Engineering and Technology, Dhaka, Bangladesh\\
\IEEEauthorrefmark{2}Department of EEE,
Ahsanullah University of Science and Technology, Dhaka, Bangladesh\\
%Dhaka, Bangladesh\\
\IEEEauthorrefmark{3}Department of CSE, University of South Florida, Florida, USA\\
Email: \IEEEauthorrefmark{1}1018052067@grad.cse.buet.ac.bd,
\IEEEauthorrefmark{2}adnan.eee@aust.edu, \IEEEauthorrefmark{3}sriramc@usf.edu,
\IEEEauthorrefmark{4}alim\_razi@cse.buet.ac.bd}}
%\end{comment}
\vspace{-.3in}

\begin{comment}
\author{\IEEEauthorblockN{1\textsuperscript{st} Saiful Islam Salim}
\IEEEauthorblockA{\textit{Department of Computer Science and Engineering(of Aff.)} \\
\textit{Bangladesh University of Engineering and Technology (of Aff.)}\\
Dhaka, Bangladesh \\
1018052067@grad.cse.buet.ac.bd}
\and
\IEEEauthorblockN{2\textsuperscript{nd} Adnan Quaium}
\IEEEauthorblockA{\textit{dept. name of organization (of Aff.)} \\
\textit{Ahsanullah University of Science and Technology (of Aff.)}\\
Dhaka, Bangladesh \\
adnan.eee@aust.edu}
\and
\IEEEauthorblockN{3\textsuperscript{rd} A. B. M. Alim Al Islam}
\IEEEauthorblockA{\textit{Department of Computer Science and Engineering (of Aff.)} \\
\textit{Bangladesh University of Engineering and Technology (of Aff.)}\\
Dhaka, Bangladesh \\
alim_razi@cse.buet.ac.bd}
}
\end{comment}

\maketitle

\begin{abstract}
Securing information transmission is critical today. However, with rapidly developing powerful quantum technologies, conventional cryptography techniques are becoming more prone to attacks each day. New techniques in the realm of quantum cryptography to preserve security against powerful attacks are slowly emerging. What is important though now is the fidelity of the cryptography, because security with massive processing power is not worth much if it is not correct. Focusing on this issue, we propose a method to enhance the fidelity of quantum cryptography using maximally entangled qubit pairs. For doing so, we created a graph state along a path consisting of all the qubits of \textit{ibmqx4} and \textit{ibmq\_16\_melbourne} respectively and we measure the strength of the entanglement using negativity measurement of the qubit pairs. Then, using the qubits with maximal entanglement, we send the modified encryption key to the receiver. The key is modified by permutation and superdense coding before transmission. The receiver reverts the process and gets the actual key. We carried out the complete experiment in the IBM Quantum Experience project. Our result shows a $15\%$ to $20\%$ higher fidelity of encryption and decryption than a random selection of qubits.
\end{abstract}


\section{Introduction}

Information security has always been a critical component of digital communications, and the need for security is only increasing. Recent advances, such as quantum computers, threaten many existing public-key cryptography systems (RSA \cite{b1, b2}, ElGamal \cite{b3}, ECC \cite{b4}, and so on). To resist the threats of quantum computing, new cryptosystems based on quantum technology, i.e., quantum cryptography, are already being explored. Quantum cryptography, a combination of quantum mechanics and classical cryptography, is an important branch of cryptography today. Although quantum cryptography is still in its infancy, its challenges to the security of conventional cryptosystems cannot be ignored. A more important concern in this regard is that the current state of quantum cryptography still falls behind in achieving high fidelity.

Therefore, in this paper, we focus on enhancing the fidelity of quantum cryptography using the maximally entangled qubits. Here, we create graph states consisting of all the qubits of \textit{ibmqx4} and \textit{ibmq\_16\_melbourne} and perform full quantum state tomography on all groups of $4$ connected qubits on the path to produce highly entangled states. Then, we use the maximally entangled qubit pairs to transmit the encryption key to the receiver. The encrypted message is sent through a conventional communication channel. As a result, we are able to demonstrate higher fidelity of quantum cryptography than the classical approach of a random selection of qubits.

Based on our work, we make the following set of contributions in this paper:
\begin{itemize}
\item We propose a new technique of quantum encryption where the encryption key is transmitted to the receiver through the quantum channel, whereas, the encrypted message is transmitted through the conventional communication channel.
\item We simulate the proposed technique in IBM Quantum Experience. 
\item We measure the fidelity of our proposed technique, which is $15\%$ to $20\%$ higher than the classical process of a random selection of qubits.
\end{itemize}

The rest of this paper is organized as follows. Section 2 introduces some related research studies about the fidelity of quantum cryptography. Section 3 discusses the background of quantum physics and quantum communication. Section 4 presents our proposed mechanism. Section 5 discusses the experimental setup of our research. Section 6 presents the results of our experiment. Finally, Section 7 concludes our paper.

\vspace{-.1in}
\section{Related Work}
Researchers are actively investigating the design of components and systems involved in quantum cryptography today. A notable secure communication method, that implements a cryptographic protocol involving components of quantum mechanics, is Quantum Key Distribution (QKD) \cite{b5}. Mirhosseini et al., \cite{b6} investigated that relying on the polarization of light for encoding, QKD limits the amount of information that can be sent per photon as well as confined the error rates. They also showed that multilevel QKD systems based on spatial-mode encoding can be more resilient against eavesdropping attacks in addition to having an increased information capacity. Milicevic et al., \cite{b7} introduced a quasi-cyclic code construction for multi-edge codes, that is highly suitable for hardware-accelerated decoding on a graphics processing unit (GPU). Pirandola et al., \cite{b8} designed a coherent-state network protocol to achieve remarkably high key rates at metropolitan distances. 

Cryptographers have been working on quantum-resistant algorithms and lattice-based cryptography \cite{b9}. However, the high computational complexity of these algorithms makes it challenging to implement lattice-based protocols on resource-constrained IoT devices. To address this challenge, Banerjee et al., \cite{b10} presented a lattice cryptography processor with configurable parameters, which results in a $124$K-gate reduction in the system area. Liu et al., \cite{b11} efficiently implemented crypto-systems for $8$ and $32$-bit micro-controllers. Apart from that, Ottaviani et al., \cite{b12} have shown that super-additivity of two-way Gaussian quantum cryptography enhances security performance. Kabir et al. \cite{b13} proposed a new technique of encryption, called Supercrypt, which enhances the security level by a significant margin with the help of quantum computing as well as enhances data transmission rate through exploiting the notion of Superdense Coding.

Fidelity measures in various types of quantum states and operators are also explored by  researchers. Gutoski et al., \cite{b14} introduced a definition of the fidelity function for multi-round quantum strategies, which is a generalization of the fidelity function for quantum states. They illustrate an operational interpretation of the strategy fidelity in the spirit of Uhlmann's Theorem and discuss its application to the security analysis of quantum protocols for interactive cryptographic tasks, such as bit-commitment and oblivious string transfer. Gyongyosi et al., \cite{b15} showed an effective method to compute the fidelity of quantum cloning based attacks in quantum cryptography using Delaunay tessellation. 

As we see existing studies have focused on quantum cryptography and fidelity. However, enhancing the fidelity of quantum cryptography and related analysis is still in its infancy, and yet to be focused in the literature. This paper designs a solution to enhance the fidelity of quantum cryptography using maximally entangled qubit pairs.

\section{Background of Quantum Cryptography}

Quantum computing uses the principles of quantum physics to perform operations on data. In our proposed technique, a very important step is creating a full quantum entangled state. Therefore, we discuss a few necessary basics of quantum entanglement along with the quantum cryptography and fidelity in this section.

\subsection{Quantum Computing}
Quantum computing is based on quantum bit or qubit, an analogous concept of the bit. The computation mainly deals with quantum information. A qubit is  different from a classical bit, which has a state of either $0$ or $1$. On the contrary, a qubit has a quantum state that can be a superposition of both the classical states ($0$ and $1$) at the same time. This quantum state, also known as superposition state, is a linear combination of the classical states.

The quantum state can be expressed as: $\vert \psi \rangle = \alpha \vert 0 \rangle + \beta \vert 0 \rangle $, where $\alpha$ and $\beta$ are probability amplitudes and both can be complex numbers in general. The two states $\alpha \vert 0 \rangle$ and $\alpha \vert 1 \rangle$ are called computational basis states and they form an orthonormal basis for computation in a vector space \cite{b16}. Utilizing the superposition states, the quantum computation can deal with a huge number of calculations simultaneously. 

A quantum computer is a device that performs quantum computing. A quantum computer with $400$ basic units (qubits) could, for example, simultaneously process more bits of information than the number of atoms in the universe \cite{b17}. Therefore large-scale quantum computers are theoretically able to rapidly solve certain problems than any classical computer.

As of 2020, the development of actual quantum computers is still in its infancy, but experiments have been carried out in which quantum computational operations were executed on a very small number of quantum bits. A small $20$-qubit quantum computer has been developed and is available for experiments via the IBM Quantum Experience project \cite{b18}. D-Wave Systems has been developing its own version of a quantum computer that uses quantum annealing \cite{b19}. Our experiment was carried out in the IBM Quantum Experience.

\subsection{Quantum Entanglement}
Quantum entanglement of particles is a quantum mechanical phenomenon, that describes a relationship between their fundamental properties that cannot happen by chance even though the individual objects may be spatially separated \cite{b20}. Quantum entanglement occurs when particles such as photons, electrons, molecules, etc interact physically and then become separated. This interaction properly describes each resulting member of a pair by the same quantum mechanical state. This could refer to states, such as their momentum, position, or polarisation.

In the case of two entangled particles, if one is observed to be spin-up, the other one will always be observed to be spin down and vice versa. However, according to quantum mechanics, it is impossible to predict, which set of measurements will be observed.

For example, let Alice and Bob be two observers for system $A$ and system $B$ respectively. In the entangled state, if Alice measures the eigenbasis of $A$, there are two possible equally probable outcomes:
\begin{enumerate}
    \item Alice measures $0$, and the state of the system collapses to $\vert 0 \rangle_A \otimes \vert 1 \rangle_B$. So, any subsequent measurement performed by Bob will always return $1$.
    \item Alice measures $1$, and the state of the system collapses to $ \vert 1 \rangle_A \otimes \vert 0 \rangle_B$ and Bob's measurement will return $0$ with certainty.
\end{enumerate}
Thus, system B is altered depending on Alice's measurement on system A. This remains true, even if the systems A and B are spatially separated.

\subsection{Quantum Cryptography}
Quantum cryptography exploits quantum mechanical properties to perform cryptographic tasks. Quantum cryptography allows the completion of various cryptographic tasks, that are proven or conjectured to be impossible using only classical (i.e., non-quantum) computation \cite{b21}. It is a special method of securely communicating a private key, from one party to another for use in one-time pad encryption \cite{b22}.

The correct selection of bases for measurement of qubits is fundamental to quantum cryptography. A sender encodes the one-time pads in strings of qubits by performing some quantum operations using particular bases and then sends it over a public quantum channel. Only the sender knows the actual bases of the performed quantum operations. Therefore the receiver cannot distinguish all original states of the qubits.

\subsection{Fidelity}

Fidelity is a measure of the distance between two quantum states. Fidelity can be explained by assessing how good the source or quantum state preparation is. This can be done by comparing the state of the measured value with the ideal value. Fidelity, denoted by $F$, is by definition $F\in[0, 1]$. Here, $F = 1$ means that two states are identical and $F = 0$ means they are as different as physically distinct possible. 

If the objective is to create pairs of perfectly entangled particles, and if there are extra stray (non-entangled) photons that were measured in an ensemble of prepared bi-photons, then the ensemble average state will be different from the expected perfectly entangled bi-photon pairs. In this case, how close the result is to the perfectly entangled bi-photon pairs is given by fidelity $F$.

In quantum cryptography, fidelity measures the correctness of the information. Higher fidelity indicates more correctness of the information. Therefore, achieving high fidelity is of critical  importance in quantum cryptography.

\section{Proposed Mechanism}

Our proposed mechanism involves, creating a full field entangled state of qubits \cite{b23}. Then highly entangled states, namely the graph states \cite{b24} are produced using optimized low-depth circuits that are tailored to the universal gate set. We detect full entanglement of all the qubits, using an entanglement criterion based on reduced density matrices. Qubits are fully entangled in the sense that, the state involves all physical qubits and is inseparable to any fixed partition.

For a set of vertices $V$ and a set of edges $E$, the graph state that corresponds to $G(V, E)$ is the unique common eigenvector (of eigenvalue $1$) of the set of independent commuting operators,

\vspace{-.2in}
\begin{equation}\label{eigenvectorKa}
    K_a = X^aZ^{N_a} = X^a \prod_{b\in N_a}Z^b
\end{equation}

where, $X$ and $Z$ denote the Pauli operators,  the eigenvalues to $K_a$ are $+1$ for all $a \in V$ , and $N_a$ denotes the set of neighbor vertices of $a$ in $G$ \cite{b25}. A $n$-qubit graph state can be prepared by the following steps:

\begin{enumerate}
    \item Initialize the state to $\vert{+} \rangle^{\otimes n}$ by applying $n$ Hadamard gates to $\vert{0} \rangle^{\otimes n}$
    \item For every $(a, b) \in E$, apply a control-$Z$ gate on qubits $a$ and $b$; the order can be arbitrary
\end{enumerate}

According to one of the most widely used criteria, partial transpose criterion \cite{b26, b27, b28}, a bipartite state $\rho_{AB}$ on the Hilbert space $\mathcal{H} = \mathcal{H}_A \otimes \mathcal{H}_B$ is said to be separable if $\rho_{AB}$ can be written as,

\vspace{-.2in}
\begin{equation}
    \rho_{AB} = \sum_i p_i \rho^i_A \otimes \rho^i_B
\end{equation}

where, $\rho^i_A$ and $\rho^i_B$ are quantum states of the system $A$ and $B$, respectively, with  the  positive  weights $p$ to be $p_i \ge 0$ and $\sum_ip_i = 1$. Otherwise $\rho_{AB}$ is entangled. For a state $\rho$ of a many-body system, for any fixed bipartition $AB$ of the system, if $\rho$ is entangled to the partition $AB$, then the entanglement of the many-body state $\rho$ can also be examined via its subsystems. That is, if the subsystems are all entangled, the whole system must be also entangled.

To be more specific, consider a 4-qubit subsystem $\rho_{A,B,C,D}$ in an $n$ qubit system -

\vspace{-.2in}
\begin{equation}\label{eigenvectorKa}
    \rho_{A,B,C,D} = \frac{1}{4}(I + Z_AX_BZ_C)(I + Z_BX_CZ_D).
\end{equation}

\begin{figure}[!htbp]
\centerline{\includegraphics[width=0.3\textwidth]{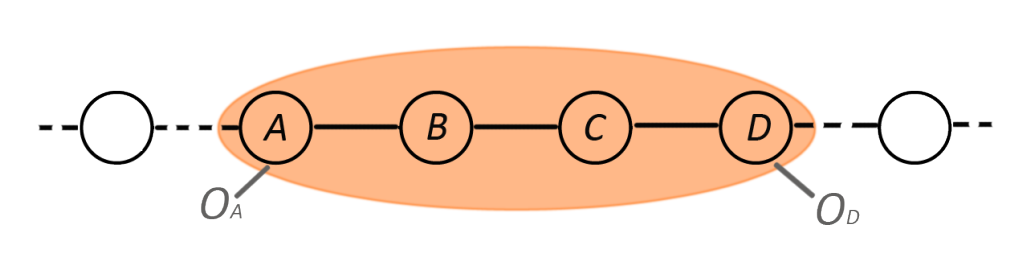}}
\caption{A four-qubit subsystem that forms a chain \cite{b23}}
\label{fig:4_qubit_subsystem}
\end{figure}

Due to Eq. \ref{eigenvectorKa}, for a ring graph state, each $4$-qubit density matrix of neighboring four qubits, as illustrated in Fig. \ref{fig:4_qubit_subsystem} is given by \cite{b23}

\vspace{-.2in}
\vspace{-.1in}

Now, to calculate the negativity of the resulting 2-qubit subsystem, local operations of  $O_A=\frac{Z_A+I}{2}$ and $O_D=\frac{Z_D+I}{2}$ need to be applied for each $4$-qubit density matrix. For example, if $(q_5, q_6, q_7, q_8)$ is chosen as our subsystem; after applying $O_A$ and $O_D$ to $q_5$ and $q_8$ respectively, $q_5$ and $q_8$ will be traced out, and the negativity of the remaining subsystem, $(q_6, q_7)$ will be measured. The reason to choose $O_A = \frac{Z_A+I}{2}$ and $O_D = \frac{Z_D +I}{2}$ is discussed below. 

If $\rho$ is graph state, and the $4$-qubit subsystem corresponds to $4$ vertices that form a chain in the graph, then the resulting $2$-qubit state is a maximally entangled state

\begin{equation}
    \vert \phi \rangle = \frac{1}{\sqrt{2}}(\vert{0} \rangle \vert {+} \rangle + \vert {1} \rangle \vert {-} \rangle).
\end{equation}

Now, two local operations $O_A$ and $O_D$ will be performed on qubit $A$ and $D$ respectively, and then, the reduced density matrix of qubit $B$ and $C$ is obtained by tracing out qubit $A$ and $D$. The reduced density matrix for qubits $B$ and $C$ will be as follows \cite{b23}

\vspace{-.2in}
\begin{equation}
    \rho_{B,C}^{'} = tr_{A,D}\Bigg(\frac{O_AO_D\rho_{A,B,C,D}O_D^{\dagger}O_A^{\dagger}}{tr\Big(O_AO_D\rho_{A,B,C,D}O_D^{\dagger}O_A^{\dagger}\Big)} \Bigg).
\end{equation}

From this, the entanglement of $\rho_{B, C}^{'}$ can be determined by using entanglement monotones such as negativity, which has non-zero values in the $2$ qubit case, if and only if the system is entangled \cite{b26, b28}. Therefore if $\rho_{B, C}^{'}$ is entangled, there is no separation with qubit $B$ and $C$ on different sides in the original system. This means, the qubit $B$ and $C$ must be on the same side for the original system to be biseparable concerning a fixed partition. The pair with maximum negativity is thus maximally entangled.

\begin{figure}[!htbp]
\centerline{\includegraphics[width=0.38\textwidth]{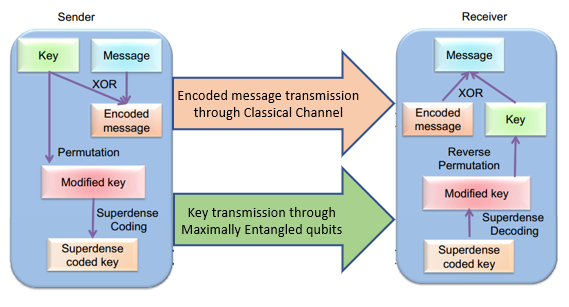}}
\caption{Block diagram of Encryption-Decryption using Maximally Entangled Qubit pairs \cite{b13}}
\label{fig:encryption_decryption}
\end{figure}

After measuring the negativity (entanglement), we perform the encryption and decryption, which is adopted from the Supercrypt protocol \cite{b13}. Fig.~\ref{fig:encryption_decryption} depicts the whole process of Supercrypt protocol. The protocol improves both security and data transmission rate simultaneously as demonstrated in \cite{b13} through comparing with other available classical alternatives, and confirms significant performance improvements in terms of both the network performance and average throughput of the network.

The encryption process consists of the following steps,
\begin{itemize}
    \item Alice (the sender) encodes the message with the key.
    \item She modifies the key using permutation.
    \item She applies the superdense coding for the key.
    \item She sends the encoded message through the \textit{classical channel} and the superdense coded key through the \textit{quantum channel}.
\end{itemize}

\vspace{-.1in}

\vspace{-.3in}

\begin{figure}[!htbp]
\centerline{\includegraphics[width=0.35\textwidth]{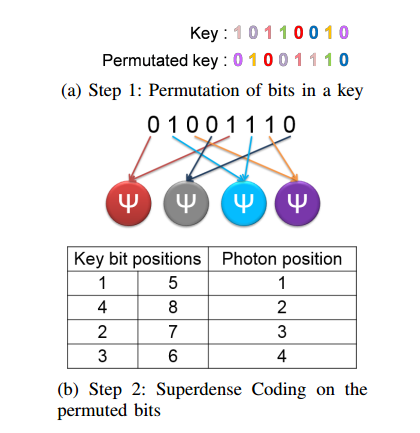}}
\caption{Encoding process in sender device \cite{b13}}
\label{fig:Encoding_process_in_sender_device}
\end{figure}
%\vspace{-.3in}

\begin{figure}[!htbp]
\centerline{\includegraphics[width=0.35\textwidth]{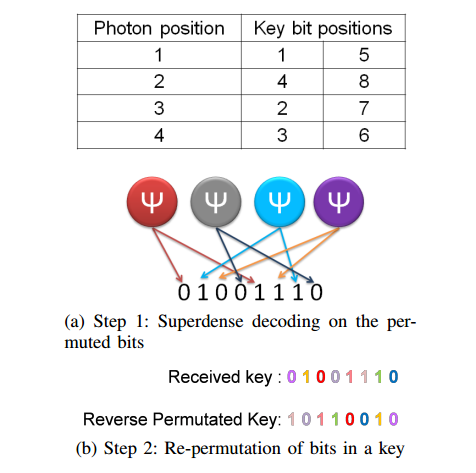}}
\caption{Decoding process in receiver device \cite{b13}}
\label{fig:Decoding_process_in_receiver_device}
\end{figure}

\vspace{-.1in}
In our proposed method, a sender needs to perform two operations to enhance the security of the transmitted key. First, a permutation operation is required on the bit sequence of the key. Second, the sender takes a pair of bits from the modified bit sequence and performs Superdense Coding on those. As a result, an $n$ bit key is encoded in $n/2$ qubits. This encoding process is elaborated in  Fig.~\ref{fig:Encoding_process_in_sender_device}.

The decryption process consists of the following steps,

\begin{itemize}
    \item Bob (the receiver) receives the encoded message from the \textit{classical channel} and the superdense coded key from the \textit{quantum channel}.
    \item He applies superdense decoding to the received key.
    \item He retrieves the modified key by applying reverse permutation.
    \item Finally he decodes the message using the key.
\end{itemize}

The decoding process confirms providing receiver an $n$ bit key from $n=2$ qubits. Subsequently, the sender performs the permutation as decided earlier and gets the original bit sequence. At the end of this phase, the sender gets the exact key or one-time pad to decrypt the message. This decoding process is elaborated in the Fig.~\ref{fig:Decoding_process_in_receiver_device}.

\section{Experimental Setup}
The IBM Quantum experience is a quantum cloud service released by IBM, which has several quantum computing devices in the backend. A full field entangled state of $5$ qubits and $14$ qubits in two machines, named \textit{ibmqx4} and \textit{ibmq\_16\_melbourne} is created respectively. Then the graph states, that correspond to $2$ rings involving $5$ qubits from \textit{ibmqx4} and $14$ qubits from \textit{ibmq\_16\_melbourne} is generated using optimized low-depth circuits that are tailored to the universal gate set. The qubit connectivity is given below,

\vspace{-.2in}
\begin{figure}[htbp]
\centerline{\includegraphics[width=0.12\textwidth]{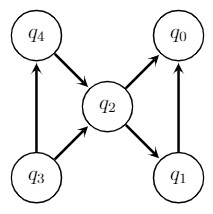}}
\caption{Qubit topology of \textit{ibmqx4} ($5$ qubits)}
\label{fig:ibmqx4_(5_qubit)}
\end{figure}

\vspace{-.3in}

\begin{figure}[htbp]
\centerline{\includegraphics[width=0.4\textwidth]{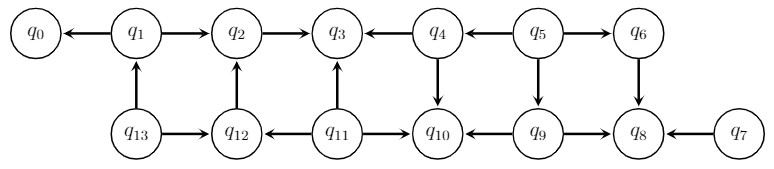}}
\caption{Qubit topology of \textit{ibmq\_16\_melbourne} ($14$ qubits)}
\label{fig:ibmq_16_melbourne_(14_qubit)}
\end{figure}

\begin{figure}[htbp]
\begin{subfigure}{\columnwidth}
\centerline{\includegraphics[width=\linewidth]{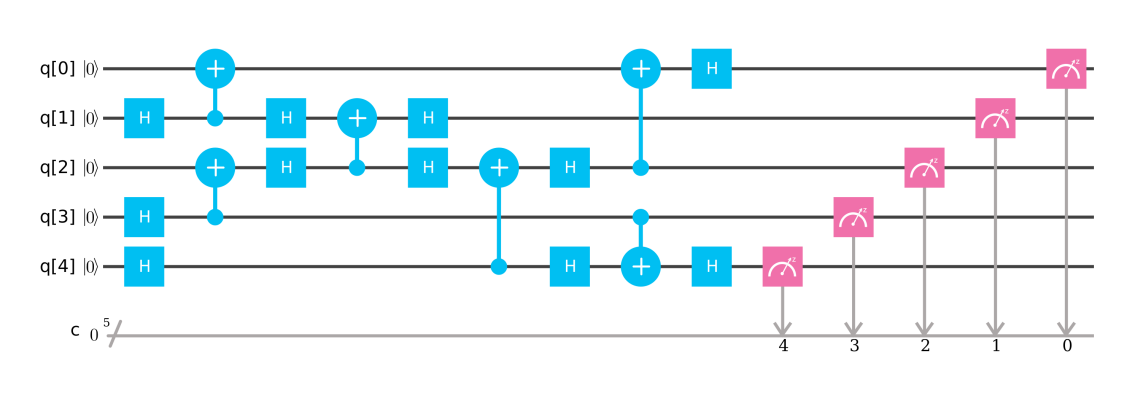}}
\vspace{-.2in}
  \caption{Full entanglement circuit in the \textit{ibmqx4} processor}
  \label{fig:entanglement_q5_optimized}
\end{subfigure}

\begin{subfigure}{\columnwidth}
\centerline{\includegraphics[width=\linewidth]{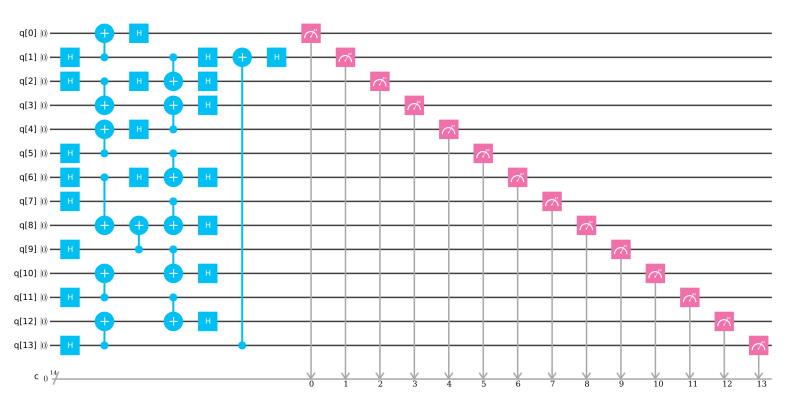}}
\vspace{-.2in}
  \caption{Full entanglement circuit in the \textit{ibmq\_16\_melbourne} processor}
  \label{fig:entanglement_q14_optimized}
\end{subfigure}
\caption{Full entanglement circuits used in our experiment}
\end{figure}

Here, Fig.~\ref{fig:entanglement_q5_optimized} and Fig. \ref{fig:entanglement_q14_optimized} show the circuit of the full entanglement that is implemented in \textit{ibmqx4} and \textit{ibmq\_16\_melbourne}.

\begin{figure}[htbp]
        \begin{subfigure}{.48\columnwidth}
         \centerline{\includegraphics[width=\linewidth]{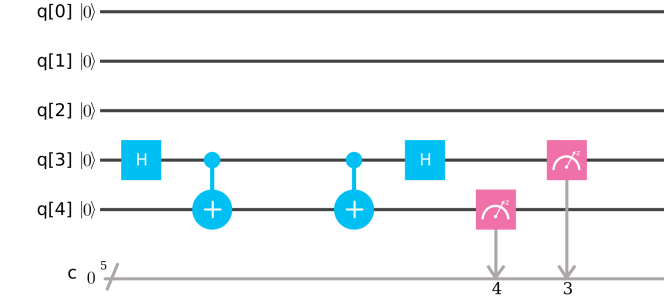}}
          \caption{Encryption-Decryption of 00}
          \label{fig:Out_00_q3_q4_new}
    \end{subfigure}\hfill
    \begin{subfigure}{.48\columnwidth}
          \centerline{\includegraphics[width=\linewidth]{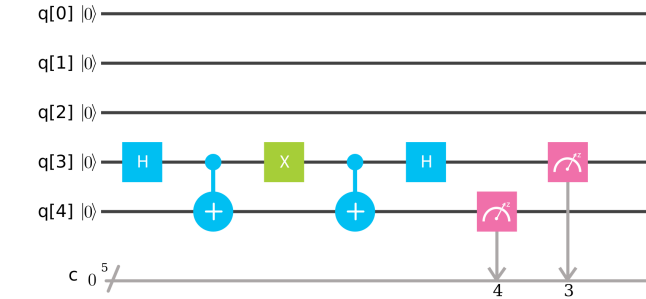}}
          \caption{Encryption-Decryption of 01}
          \label{fig:Out_01_q3_q4_new}
    \end{subfigure}\newline

    \begin{subfigure}{.48\columnwidth}
           \centerline{\includegraphics[width=\linewidth]{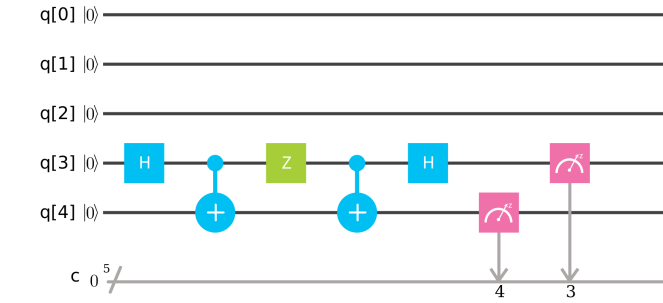}}
          \caption{Encryption-Decryption of 10}
          \label{fig:Out_10_q3_q4_new}
    \end{subfigure}\hfill
    \begin{subfigure}{.48\columnwidth}
          \centerline{\includegraphics[width=\linewidth]{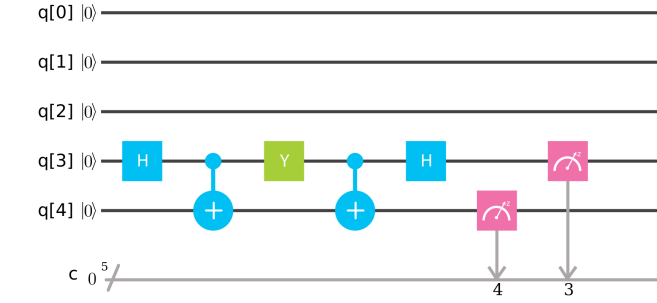}}
          \caption{Encryption-Decryption of 11}
          \label{fig:Out_11_q3_q4_new}
    \end{subfigure}
    \caption{Encryption and decryption for qubit pair $(q3,q4)$}
    \label{fig:encryption-decryption-figure}
\end{figure}

After measuring the entanglement for all the qubit pairs, the encryption and decryption process is implemented. In our experiment, two different devices to communicate with each other using encryption and decryption is not available. Therefore, an assumption is made that there is no information loss in data communication, and the encryption and decryption processes are implemented on the same device. And then, the fidelity for each predefined qubit pairs is checked. Implementation of the encryption and decryption of key in \textit{ibmqx4} for qubit pair $(q3,q4)$ is shown in Fig.~\ref{fig:encryption-decryption-figure}.

%\vspace{-.2in}

\section{Experimental Results}

For \textit{ibmqx4}, the measured qubit pairs were $(q_0, q_1)$, $(q_0, q_2)$, $(q_1, q_2)$, $(q_2, q_3)$, $(q_2, q_4)$, $(q_3, q_4)$ and for \textit{ibmq\_16\_melbourne}, the measured qubit pairs were $(q_0, q_1)$, $(q_1, q_2)$, $(q_2, q_3)$, $(q_3, q_4)$, $(q_4, q_5)$, $(q_5, q_6)$, $(q_6, q_8)$, $(q_7, q_8)$, $(q_8, q_9)$, $(q_9, q_{10})$, $(q_{10}, q_{11})$, $(q_{11}, q_{12})$, $(q_{12}, q_{13})$, $(q_{13}, q_1)$.

\begin{figure}[htbp]
        \begin{subfigure}{.49\columnwidth}
    \centerline{\includegraphics[width=\linewidth, frame]{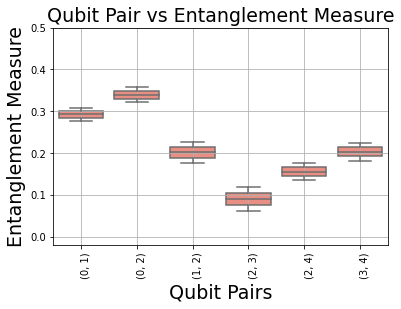}}
      \caption{ Entanglement measure in \textit{ibmqx4} processor}
      \label{fig:Qubit_pair_vs_Entanglement_measure_ibmqx4}
    \end{subfigure}\hfill
    \begin{subfigure}{.47\columnwidth}
    \centerline{\includegraphics[width=\linewidth, frame]{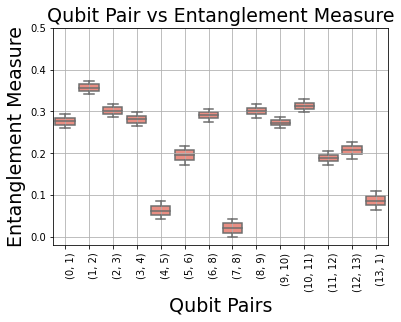}}
     \caption{Entanglement measure in \textit{ibmq\_16\_melbourne} processor}
     \label{fig:Qubit_pair_vs_Entanglement_measure_ibmq_16_melbourne}
    \end{subfigure}\newline

    \caption{Entanglement measure for each qubit pairs}
    \label{fig:Qubit_pair_vs_Entanglement_measure}
\end{figure}

In both quantum processors, the entanglement measure for each pair of qubits is calculated and the results are plotted in Fig.~\ref{fig:Qubit_pair_vs_Entanglement_measure}. The entanglement measure, i.e., Negativity, ranges between $0$, and $0.5$, where $0$ indicates no entanglement and larger values indicate more entanglement \cite{b23,b29}. We found that, the magnitude of entanglement between pairs of qubits in the \textit{ibmq\_16\_melbourne} processor surpasses the \textit{ibmqx4} processor.

\begin{figure}[htbp]
    \begin{subfigure}{0.49\columnwidth}
    \centerline{\includegraphics[width=\linewidth, frame]{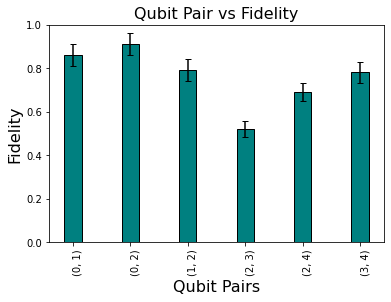}}
      \caption{Qubit vs Fidelity in \textit{ibmqx4} processor}
      \label{fig:qubit_pair_vs_fidelity_ibmqx4}
    \end{subfigure}
    \hfill
    \begin{subfigure}{0.47\columnwidth}
    \centerline{\includegraphics[width=\linewidth, frame]{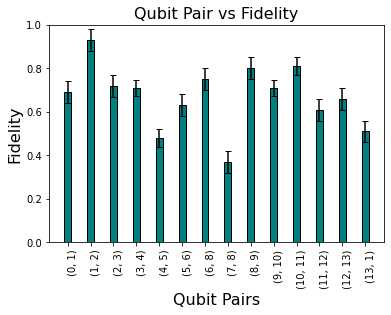}}
     \caption{Qubit vs Fidelity in \textit{ibmq\_16\_melbourne} processor}
     \label{fig:qubit_pair_vs_fidelity_ibmq_16_melbourne}
    \end{subfigure}
    \hfill

    \caption{Fidelity for each qubit pairs}
    \label{fig:qubit_pair_vs_fidelity}
\end{figure}

The fidelity for each pair of qubits for encryption and decryption in both quantum computers is measured. To determine the fidelity, total $20$ measurements are conducted, where a total of $8192$ shots were used for each measurement. The average fidelity of all the measurements is calculated, which are shown in Fig.~\ref{fig:qubit_pair_vs_fidelity}. The fidelity ranges between $0$ and $1.0$, where $0$ indicates the states of the qubits are completely different and $1.0$ indicates the states are identical. 

Now, Fig.~\ref{fig:Qubit_pair_vs_Entanglement_measure_ibmqx4} and Fig.~\ref{fig:qubit_pair_vs_fidelity_ibmqx4} clearly indicate that, qubits with the higher entanglement show higher fidelity. The same is true for Fig.~\ref{fig:Qubit_pair_vs_Entanglement_measure_ibmq_16_melbourne} and  Fig.~\ref{fig:qubit_pair_vs_fidelity_ibmq_16_melbourne}.  It is worth mentioning that, if the entanglement was not considered in this experiment, the average fidelity for a random selection of qubits would not be better. 

\begin{figure}[h]%[htbp]
\centerline{ \includegraphics[width=0.325\textwidth, frame]{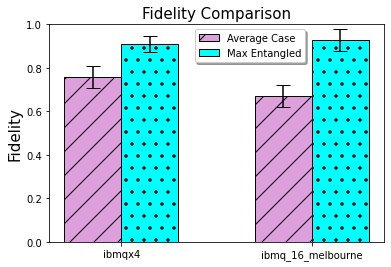}}
  \caption{Fidelity comparison between randomly selected qubits (average case) and maximally entangled qubits}
  \label{fig:average_vs_max_entangled}
\end{figure}

Considering the maximally entangled qubit pairs, we found the fidelity of the encryption and decryption process approximately $15\%$ to $20\%$ higher than the random selection of qubits. This result is shown in Fig.~\ref{fig:average_vs_max_entangled}. It can also be observed from  Fig.~\ref{fig:average_vs_max_entangled} that, the standard error of the fidelity is lower in case of the maximally entangled qubits, especially, in the \textit{ibmqx4} processor. The maximally entangled  fidelity of \textit{ibmq\_16\_melbourne} processor is slightly higher than the \textit{ibmqx4} processor.

\vspace{-.1in}

\vspace{-.2in}

\begin{comment}
\begin{figure}[htbp]
\centerline{\includegraphics[width=.4\textwidth, frame]{images/qubit_pair_vs_fidelity_ibmqx4.png}}
  \caption{Qubit vs Fidelity in \textit{ibmqx4} processor}
  \label{fig:qubit_pair_vs_fidelity_ibmqx4}
\end{figure}

\begin{figure}[htbp]
\centerline{ \includegraphics[width=.4\textwidth, frame]{images/qubit_pair_vs_fidelity_ibmq_16_melbourne.png}}
  \caption{Qubit vs Fidelity in \textit{ibmq\_16\_melbourne} processor}
  \label{fig:qubit_pair_vs_fidelity_ibmq_16_melbourne}
\end{figure}
\end{comment}

\vspace{-.3in}

\section{Conclusion and Future Work}
%\vspace{-.1in}
Recent advancements in quantum computers pose significant risks to both conventional public-key and symmetric key algorithms. As a result, new encryption techniques that could offer more security becomes necessary. 
Here, the fidelity of resulting cryptography techniques is important to ensure correctness, and the current state here has room for improvement. In this paper, we enhanced the fidelity of quantum cryptography using the notion of maximally entangled qubit pairs to transmit the super dense key to the receiver. Our experiment shows that using maximally entangled qubits, the fidelity of cryptography significantly improves by up to $20\%$ compared to
%. The result clearly indicates an approximately 15\% to 20\% higher fidelity of encryption-decryption, i.e., quantum cryptography, 
%than  
the classical method of a random selection of qubits. We plan to implement our work for a higher number of qubits in two separate devices for encryption and decryption in the near future. Furthermore, we plan to conduct a detailed correctness analysis of our proposed methodology also.

\section*{Acknowledgment}
This work was supported in part by US National Science Foundation (Grant \# 1718071). Findings and conclusions are those of the authors alone, and do not reflect views of NSF.
%\vspace{-.1in}

\bibliographystyle{IEEEtran}
\bibliography{bibliography}

\begin{comment}

\end{comment}

\end{document}